\documentclass[a4paper]{jpconf}
\usepackage{graphicx}
\begin{document}
\title{Cosmic rays from multi-wavelength observations of the Galactic diffuse emission}

\author{Elena Orlando}

\address{Hansen Experimental Physics Laboratory and Kavli Institute for Particle Astrophysics and Cosmology,
        Stanford University, Stanford, CA 94305, U.S.A.}

\ead{eorlando@stanford.edu}

\begin{abstract}
Cosmic rays (CRs) generate diffuse emission while interacting with the Galactic magnetic field (B-field), the interstellar gas and the radiation field. This diffuse emission extends from radio, microwaves, through X-rays, to high-energy gamma rays. Diffuse emission has considerably increased the interest of the astrophysical community due to recent detailed observations by Planck, Fermi-LAT, and by very-high energy Cherenkov telescopes. 
Observations of this emission and comparison with detailed predictions are used to gain information on the properties of CRs, such as their density, spectra, distribution and propagation in the Galaxy. Unfortunately disentangling and characterizing this diffuse emission strongly depends on uncertainties in the knowledge of unresolved sources, gas, radiation fields, and B-fields, other than CRs throughout the Galaxy. We report here on recent multiwavelength observations of the Galactic diffuse emission, and discuss the diffuse emission produced by CRs and its model uncertainties, comparing observations with predictions. The importance for forthcoming telescopes, especially for the Square Kilometre Array telescope (SKA) and the Cherenkov Telescope Array (CTA), and for future missions at MeV energies is also addressed.
\end{abstract}

\section{Introduction}
The interest on the Galactic diffuse emission has remarkably increased thanks to the recent high-quality observations by Planck, Fermi-LAT, H.E.S.S. and VERITAS. In fact diffuse emission from our Galaxy is a common element from radio, microwaves, through X-rays, to high-energy gamma rays, and it is more pronounced in the direction of the Galactic center. 
CRs generate non-thermal Galactic diffuse emission components while propagating on the Galaxy and interacting with the B-field, the gas and the radiation field. A schematic cartoon of these components is shown in Figure 1. 

\section{Observations of the diffuse emission}
\subsection{Radio}
Radio surveys, especially the 408 MHz Haslam map  \cite{haslam81}, are representative of the non-thermal emission from the Galaxy due by synchrotron radiation of CR electrons and positrons on B-fields. The contribution by unresolved radio sources may also be significant.
\subsection{Microwaves}
At higher frequencies, from tens of GHz, many radiation mechanisms are generating diffuse emission. These are: synchrotron emission; free-free emission from interactions of thermal electrons with ionized gas; thermal dust radiation; and anomalous microwave emission believed to be from spinning dust from small rapidly grains. Disentangling these components in the {\it WMAP} and {\it Planck} band requires various assumptions or a priori information such as data-driven templates using ancillary data or by approximating their spectral properties (see \cite{bennett, Ingunn, Paddy}). The high-quality {\it Planck} component maps recently released \cite{Ingunn} were combined with the complementary 9-year {\it WMAP} temperature maps and the 408\,MHz map \cite{haslam81, 408new}, providing an excellent fit to the data. As baseline model we used the synchrotron spectral index from the best model by \cite{SOJ2011, O&S2013} to allow a better separation of the synchrotron component and consequently to gain information on the B-field \cite{Tess}. Nevertheless, there is likely to be a degeneracies between the various low-frequency components, especially between anomalous microwave emission and synchrotron, as discussed in detail in \cite{Paddy}.  In polarization the Galactic components are synchrotron emission and thermal dust only. The new-released Planck synchrotron polarization map shows many loops and spurs. We revealed  \cite{Paddy} a harder synchrotron spectral index in the filament around the northern Fermi Bubble. This suggests that electrons there have harder spectrum, and it might imply that the filament is connected to the Bubble. However the same does not happen in the southern Bubble. This seems to suggest that the Fermi Bubbles and the filament are two different structures at different distances.
\begin{figure}
\includegraphics[width=16pc]{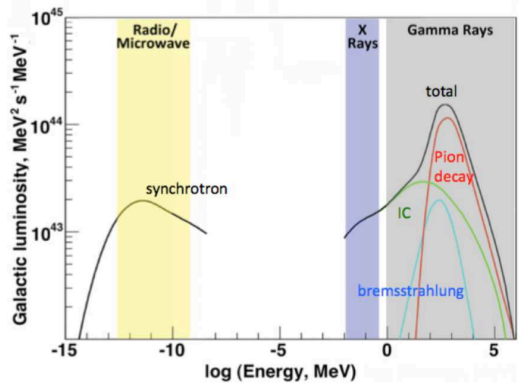}\hspace{1pc}%
\begin{minipage}[b]{18pc}\caption{\label{label} Major emission mechanisms (left to right): - Radio and Microwave (yellow area): Synchrotron by CR leptons spiraling in the B-field. - X-rays (violet area): inverse Compton (IC) by CR leptons on the cosmic microwave background, infrared and optical photons. - Gamma rays (gray area): IC, pion decay due by CR hadronic interactions on the gas, and bremstrahlung of CR leptons on the gas. Model example from \cite{strong2010}.}
\end{minipage}
\end{figure}
\subsection{High energies}
At the opposite side of the electromagnetic spectrum it is believed that $\sim$50\% of the photons seen by Fermi-LAT has Galactic diffuse emission origin \cite{diffuse2}. This diffuse emission contains unresolved sources and emission due to CRs.  Some of it may be possibly due to dark matter. The sky above 100 MeV is dominated by emission produced by CRs interaction on gas and interstellar photon fields via $\pi^0$-decay, IC, and bremsstrahlung. The situation below 100 MeV has been barely explored (see Fig 2). Extrapolations of present models to such low energies predict IC and bremsstrahlung to be the major mechanisms of CR-induced emission. 
\begin{figure}[h!]
\includegraphics[width=12pc]{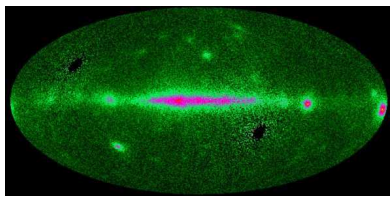}\hspace{1pc}%
\begin{minipage}[b]{22pc}\caption{\label{label} Fermi-LAT intensity map below 80 MeV for  5-year Pass 8 data applying extreme cuts to exclude solar, lunar and Earth's limb emission (Orlando for the Fermi-LAT coll., 2014 Nagoya Fermi Symposium). Here the emission is believed to be mostly of leptonic origin.}
\end{minipage}
\end{figure}
Disentangling the different components is challenging and is usually done in a model-dependent approach. An intensive studies of the Fermi Bubbles \cite{Su} was recently presented \cite{Bubbles}, however the emission mechanism is still not individuated. In the last years, many authors found an excess of diffuse emission in the inner Galaxy with respect to usual interstellar models. There is not consensus yet whether this is given by star formation, unresolved sources, enhanced CRs or dark matter annihilation.
\subsection{Very-high energies} 
At even higher energies H.E.S.S. completed the survey of the Galactic plane \cite{hess_icrc}, observing large-scale emission interpreted as a mix of truly diffuse emission and unresolved sources. The updated analysis of the Galactic center ridge within the central 3$^\circ$ in longitude with the more extended observation by H.E.S.S. above 300 GeV confirmed the presence of diffuse emission in this region, after subtracting the sources. The emission follows the gas CS template up to projected distance of 1$^\circ$ or 150 pc, supposing an hadronic origin due to collisions of multi-TeV CRs with the dense clouds of interstellar gas present in this region. In addition they found a diffuse component that do not correlate with gas tracers with a larger extension, and which origin is still unclear. A diffuse gas component, not well accounted for by dense gas tracers such as CS, might be responsible for this emission. A number of unresolved sources can probably contribute as well. They also found an additional feature in the central 30 pc that could be a signature of a stationary CR source accelerated by a SMBH in the Cloud Molecular Zone. At even higher energies recently also VERITAS \cite{Veritas} observed a residual emission above 2 TeV in the inner degree around the Galactic center after subtracting the sources, which contours overlaps with the ones seen by H.E.S.S..
\section{Our goals and method}
Understanding the physics behind the observed diffuse emission and gaining information on CRs need multiwavelength observations and possibly the use of propagation models. Our approach on extracting information on interstellar CRs and their distribution, spectral indexes of the injected particles, and propagation parameters is by looking simultaneously at the diffuse emission in radio, microwaves and gamma rays accounting for CR direct measurements and advanced modeling. We use the GALPROP code\footnote{http://galprop.stanford.edu/} \cite{moskalenko98, strong2007} that solves the transport equation accounting for energy losses, diffusion, acceleration, convection, fragmentation, radioactive decay for all CR species. Modeling the induced CR diffuse emission needs a good knowledge of the B-field and the interstellar medium. Unfortunately these quantities are affected by big uncertainties. We discuss here the most significant ones at the various wavelength. 
\begin{figure}
\begin{minipage}{18pc}
\includegraphics[width=15pc]{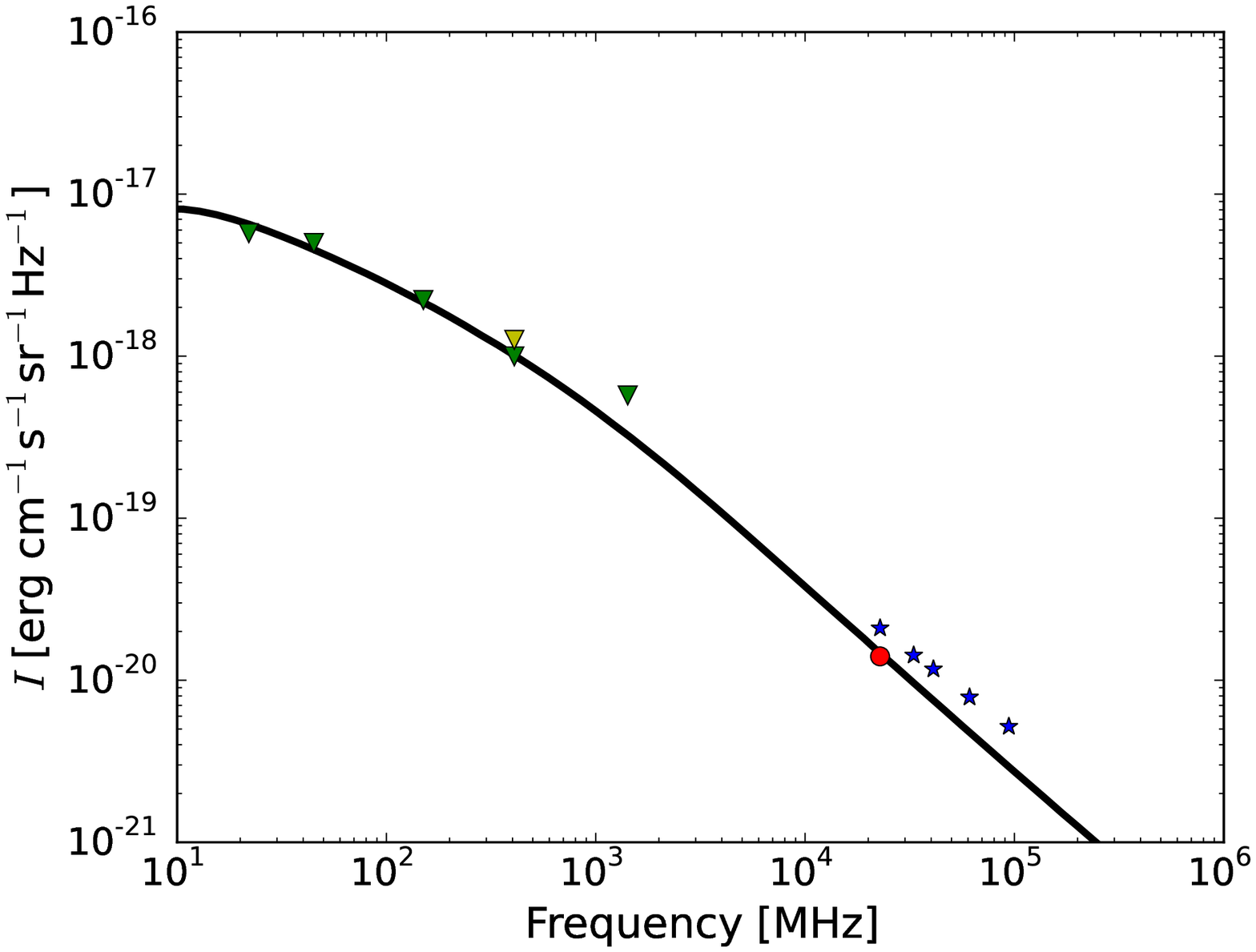}
\caption{\label{label}Synchrotron spectrum ({\it black line}) from high latitudes ($10^{\circ}<|b|<45^{\circ}$, all longitudes) for a propagation model calculated with the CRs as measured by \cite{Pamela}.  Data are:  9-year {\it WMAP} synchrotron maps ({\it blue stars}) \cite{bennett}; {\it Planck} synchrotron maps scaled to 23\,GHz \cite{Paddy} ({\it red point}); radio surveys described in \cite{SOJ2011} ({\it green triangles}); 408\,MHz from \cite{haslam81} with 3.7\,K offset ({\it yellow triangle}) and reprocessed 408\,MHz from \cite{408new} with 8.9\,K offset ({\it green triangle}).}
\end{minipage}\hspace{1pc}%
\begin{minipage}{17pc}
\includegraphics[width=15pc]{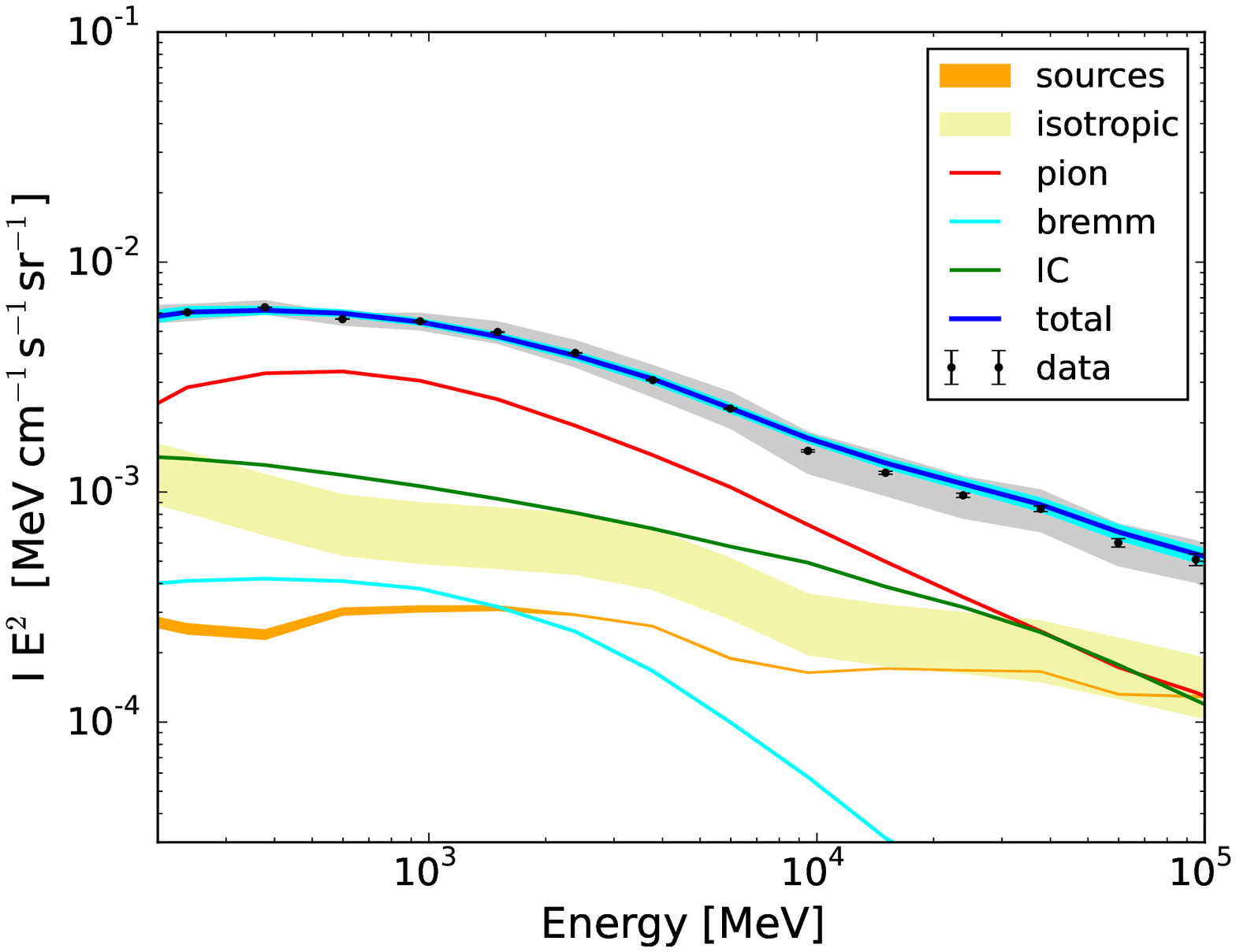}
\caption{\label{label}Gamma-ray spectral components compared with the Fermi-LAT data from \cite{diffuse2} for intermediate latitudes ($10^{\circ}<|b|<20^{\circ}$, all longitudes), using the plain diffusive propagation model and the lepton spectrum as used in Fig 3. Data include statistical ({\it grey area}) and systematic errors ({\it black bars}). Here components are not fitted to gamma-ray data, so uncertainties in the gas and ISRF are not accounted for. }
\end{minipage} 
\end{figure}
\section{Uncertainties in the modeling}
The first uncertainty comes from the {\it CR source distribution} in the Galaxy. In fact, CR source properties and position are not firmly constraint, especially far away from our local position. Then, the {\it B-field}, which is responsible of energy losses and the synchrotron diffuse emission is still unknown, despite intensive studies. Polarized synchrotron emission traces the regular and ordered components of the B-field only, while the unpolarized synchrotron emission traces the total B-field (random plus ordered component). The topology of the ordered and random components, the presence of a toroidal component in the halo, of an additional striated component not traced with rotation measurements, and the intensity of the various components throughout the Galaxy are still not clear.
Also the {\it interstellar gas}, which is responsible for the hadronic emission in gamma rays, provides uncertainties to the modeling. The gas is made by hydrogen and helium. Hydrogen is in the atomic, molecular and ionized form. The gas is better traced by dust emission, but the later does not contain velocity information. From here the need of using emission lines. The atomic hydrogen is traced by the 21 cm emission line corrected from optical depth, usually assuming the same spin temperature along the line of sight, which is an approximation. Molecular hydrogen is traced by the observable CO emission line, but the conversion between intensity of CO emission line and the column density of molecular hydrogen is not known and can vary in the Galaxy. Moreover, velocity information of the emission lines does not work in the direction of the Galactic center. There is also evidence of the existence of dark neutral medium, which is not traced by HI, nor by CO, for which the extinction maps are used.
For modeling the IC component we need also a good model of the {\it interstellar radiation field} (ISRF), however, again, stellar distribution and properties are not precisely known. Note that all these measured or observed quantities are integrated along the line-of-sight, making the modeling even more challenging.\\ 
As said above, the interstellar emission encloses information on CRs and their {\it propagation parameters}, which are at the moment still unknown, as example: CR spectra throughout the Galaxy and at injection, propagation halo size, solar modulation (that makes the lower energy CR spectrum impossible to directly measure), diffusion coefficient (that can vary in the Galaxy), presence of convection and/or reacelleration.

\section{Some recent results}
A detailed study of the CR-induced diffuse emission from 200 MeV to 10 GeV was performed in 2012 \cite{diffuse2}, where many CR reacceleration models (various gas, ISRF and CR source distributions, and propagation halo size) were fitted to the Fermi-LAT data. Even though all the models described quite well the data, two issues came up: 1) many model-dependent structures (eg. Fermi bubbles, Loop I, outer Galaxy, inner Galaxy) showed up as excess over the model adopted; 2) none (or a set of) best-fit model could be individuated for the whole sky. The first issue was due to the non-inclusion of these structures in the models, or to the need for different or more sophisticated models (the same topology of reacceleration models was tested). The second issue instead was due to the highly degenerate model parameters. In addition, many authors are continuously confirming the inner Galaxy excess with respect to standard models. Weather this is attributed to unresolved sources, modified CR source distributions, different gas or ISRF model, enhanced CRs or dark matter is still an open question. 
Parallel to our effort in gamma rays, we modeled in radio and microwaves the synchrotron brightness temperature and polarization \cite{O&S2013} for different models (with/without reacceleration, different CR source distributions, various 3D turbulent and ordered B-field models, propagation halo size). A best model was individuated, but it was not well reproducing the data for the whole sky. In fact the degeneracy between the topology of B-field and the spatial distribution of CRe is unsolved. 
More recently we accounted for the PAMELA CR lepton measurements \cite{Pamela}. The derived synchrotron emission was calculated and its spectrum was compared with radio surveys, the reprocessed 408\,MHz map \cite{408new} and recent observations from the 9-year {\it WMAP} and four-year {\it Planck} synchrotron temperature maps. The gamma-ray diffuse emission was calculated and compared to Fermi-LAT data. We report here on the spectral properties only. We have found a local interstellar spectrum of electron plus positrons that fit PAMELA direct measurements after correcting for modulation, and simultaneously it fits the synchrotron spectral index of radio and microwave synchrotron maps (Fig 3) and the gamma-ray spectrum seen by Fermi-LAT (Fig 4). Propagation parameters are also derived for that model. More details on the method and results are in Orlando et al. ICRC 2015 Conference Proceedings. Extension to lower and higher energies observed by Fermi-LAT and the spatial study at both gamma rays and radio/microwaves will be presented elsewhere. 

\section{Importance for forthcoming telescopes }
The forthcoming SKA will observe the sky over a wide range of frequencies and it will be about fifty times more sensitive than any other radio instrument. Thanks to the angular resolution and survey speed capability of SKA the continuum surveys of the sky will be very detailed and deep, and the importance of modeling the diffuse emission is clear. With our approach we will be able to increase our knowledge of CRs, B-field, and to better model the synchrotron emission. 
At the opposite side of the spectrum, at high energies, CTA will provide images of the sky with unprecedented angular resolution and sensitivity, revealing a lot on CRs and their site of acceleration. The recent diffuse emission observed in the Galactic ridge with imaging Cherenkov telescopes will be understood and modeled. This will be of benefit also for planning future mission and its scientific outcome at MeV energies, for example with ASTROGAM, which will observe a major diffuse component. 

\vspace*{0.5cm} 
\footnotesize{{\bf Acknowledgment:}{
 E.O. acknowledges support via NASA Grant No. NNX15AU79G.}}

\end{document}